\documentclass[superscriptaddress,aps,pra,amssymb,amsfonts]{revtex4}

\usepackage{amsmath,amsthm}
\usepackage{amsfonts}
\usepackage{hyperref}

\usepackage{graphicx,color,graphics}

\newcommand{\beq}{\begin{equation}}
\newcommand{\eeq}{\end{equation}}
\newcommand{\bdm}{\begin{displaymath}}
\newcommand{\edm}{\end{displaymath}}
\newcommand{\bea}{\begin{eqnarray}}
\newcommand{\eea}{\end{eqnarray}}
\newcommand{\nn}{\nonumber}

\newcommand{\benum}{\begin{enumerate}}
\newcommand{\eenum}{\end{enumerate}}
\newcommand{\bit}{\begin{itemize}}
\newcommand{\eit}{\end{itemize}}
\newcommand{\bdes}{\begin{description}}
\newcommand{\edes}{\end{description}}

\newcommand{\binomial}[2]{\left(\begin{array}{c}#1\\#2\end{array}\right)}

\newcommand{\si}{\sigma}

\newcommand{\s}[2]{\sigma_{#1}^{(#2)}}

\newcommand{\Ls}{L_{\sigma}}
\newcommand{\2}{\frac{1}{2}}
\newcommand{\4}{\frac{1}{4}}

\newcommand{\cQ}{\cal{Q}}

\newcommand{\PPD}{\check{P}_{ q, q_3 , \sigma_3}^{(Z)}}

\begin{document}

\title{Generalized Dicke States}

\author{S. Hartmann}
\affiliation{Munich Center for Mathematical Philosophy, LMU Munich, Geschwister-Scholl-Platz 1, 80539 Munich, Germany}
\email {S.Hartmann@lmu.de}

\date{\today}
\begin{abstract}
 Quantum master equations are an important tool in quantum optics and quantum information theory. For systems comprising a small to medium number of atoms (or qubits), the non-truncated equations are usually solved numerically. In this paper, we present a group-theoretical superoperator method that helps solving these equations. To do so, we exploit the $SU(4)$-symmetry of the respective Lindblad operator and construct basis states that generalize the well-known Dicke states. This allows us to solve various problems analytically and to considerably reduce the complexity of problems that can only be solved numerically. Finally, we present  three examples that illustrate the proposed method. 
\end{abstract}

\maketitle



\pagestyle{myheadings}
\thispagestyle{plain}
\markboth{S. HARTMANN }{Generalized Dicke States}


\section{Introduction}

Quantum master equations have been used to address a variety of problems from quantum optics, such as the description of laser systems and the phenomenon of resonance fluorescence  \cite{BP, Ca1, Ca2}. In quantum information theory, they are used to study the decoherence of entangled quantum states \cite{NC}. For a finite number $Z$ of 2-level atoms (or qubits), quantum master equations involve non-unitary terms of the Lindblad form:
\bea \label{eq:damp1}
\Ls^{(Z)} P =
&-& \frac{B}{2} (1 - s) \  \sum_{i = 1}^{Z}  \
[\s{+}{i} \s{-}{i}P + P \ \s{+}{i} \s{-}{i} - 2 \s{-}{i} P \s{+}{i}]  \nn  \\
&-& \frac{B}{2} s \ \sum_{i = 1}^{Z} \
[\s{-}{i} \s{+}{i}P + P \ \s{-}{i} \s{+}{i} - 2 \s{+}{i} P \s{-}{i}]   \\
&-& \frac{2 C - B}{4} \  \sum_{i = 1}^{Z} \ [P - \s{3}{i} P \  \s{3}{i}]  \nn 
\eea
Here $P$ is the density operator of the system under consideration, and $\s{\pm}{i}$ and $\s{3}{i}$ are the usual Pauli matrices, acting on atom $i$ whose states are $\vert 1>$ or  $\vert 0>$, representing, respectively, the atom being in the excited state or in the ground state.  $B$ and $C$ are decay constants and $s$ is the pumping parameter \cite{BE}. It varies from $s=0$ for pure damping to $s=1$ for full laser action.

To solve the corresponding quantum master equations, three approaches have been taken: First, one focuses on the case of one atom. Second, one truncates eq. (\ref{eq:damp1}) and derives semi-classical models. Third, one employs numerical simulation methods such as the quantum trajectory method  \cite{Ca2}. While the latter method is very popular, it should be noted that the numerical complexity of the problem increases exponentially with the number of atoms, and so numerical methods soon become unfeasible. To proceed also in these cases, we propose an analytical approach that exploits the symmetry of the Lindblad operator. 

We will show in this paper that Lindblad operators such as the one in eq.  (\ref{eq:damp1}) are invariant under $SU(4)$ transformations. This observation allows us to construct appropriate basis states for the density operator $P$. These basis states can be classified according to their symmetry type $Y$, which is preserved by the unitary and non-unitary dynamics of the system. Our approach has two advantages: First, being able to express the Lindblad operator in terms of the generators of $SU(4)$ allows us to solve a large class of problems analytically. Second, the observation that the dynamics preserves the symmetry type $Y$ leads to a considerable reduction of the number of required basis states for numerical calculations. For example, for  fully symmetrical states, the number of basis states is of the order $Z^3$ (see eq. (\ref{dimsym})), which has to be compared to $4^Z$ if the symmetry of the dynamics is not taken into account.

The remainder of this paper is organized as follows. Sec. \ref{s:dicke} introduces Dicke states and shows that they do not suffice to solve quantum master equations that involve non-unitary terms such as those in eq.  (\ref{eq:damp1}). Sec. \ref{s:gds} introduces generalized Dicke states and explicitly constructs fully symmetrical basis states. Sec. \ref{s:solving} shows how the proposed method can be used to solve quantum master equations. Sec.  \ref{s:apps} present  three examples that illustrate the proposed method.  Finally,  Sec. \ref{s:conclusions} suggests a number of further applications that we plan to address in future work.

\section{Dicke States} \label{s:dicke}

As will become apparent below, the new basis states that we propose in this paper generalize the well-known Dicke states  \cite{D}. To show the need for a generalization, we first introduce Dicke states and show that they do not suffice to solve quantum master equations involving terms such as those occurring in eq. (\ref{eq:damp1}). We consider a system of $Z$ 2-level atoms, of which $n_+$ atoms are in state $|1>$ and  $n_- = Z - n_+$ atoms are in state $|0>$. Using the analogy to the standard $SU(2)$ angular momentum eigenstates, Dicke states are characterized by two quantum numbers, $S$ and $S_3$. They are defined as follows:
\beq \label{dicke}
|S, S_3> = {\cal S} | \{ n_{+}, n_{-} \} >
\eeq
Here ${\cal S}$ is the symmetrization operator, $S = Z/2$ and $n_{\pm} = S \pm S_3$ with $-S \le S_3 \le S$.  As an example for a Dicke state of three atoms, we calculate
\bea
| 3/2, 1/2> &=& {\cal S} \,  | \{  2, 1 \} >  \nn \\
&=& \frac{1}{\sqrt{3}} \, ( |110> + |101>  + |011> ) \nn .
\eea

Dicke states are eigenstates of $\hat{S}^2$ (with eigenvalue $S(S+1)$) and $\hat{S}_3$ (with eigenvalue $S_3$). Raising and lowering operators ($\hat{S}_\pm$) and $\hat{S}_3$ can be defined accordingly: 
\beq
 \hat{S}_\pm = \sum_{i = 1}^{Z}  \  \s{+}{i} \quad , \quad   \hat{S}_3 = \2 \, \sum_{i = 1}^{Z}  \ \s{3}{i}
 \eeq
They satisfy the commutation relations
\beq
\lbrack  \hat{S}_+ ,  \hat{S}_- \rbrack = 2 \,  \hat{S}_3  \quad , \quad \lbrack  \hat{S}_3,  \hat{S}_\pm  \rbrack = \pm  \hat{S}_\pm  .
\eeq

Using Dicke states, it is straight forward to construct basis states for density operators for a system composed of $Z$ atoms:
\beq
P^{(D, Z)}_{M, M'} = |Z/2 , M><Z/2 , M'|, \quad {\rm for} \ M, M' = -Z/2,\dots, Z/2
\eeq
Once the number of atoms is fixed, these states are characterized by two quantum numbers, $M$ and $M'$. The basis has the dimension 
\beq \label{dimdick}
D_D(Z)= (Z+1)^2.
\eeq

Unfortunately Dicke states do not suffice to diagonalize quantum master equations for $Z$ atoms that involve non-unitary terms such as those in eq. (\ref{eq:damp1}). To show this, we first note that the first two terms in eq. (\ref{eq:damp1}) work fine as $\sum_{i = 1}^{Z} \, \s{+}{i} \s{-}{i}  = Z/2 + \hat{S}_3$ and $\sum_{i = 1}^{Z} \, \s{-}{i} \s{+}{i}  = Z/2 - \hat{S}_3$. However, calculating  $\sum_{i=1}^3 \s{-}{i}  P \s{+}{i}$ for, say, $P^{(D, 3)}_{3/3,3/2}  = | 3/2 , 3/2> <3/2 , 3/2 | = |111><111|$ yields
\beq \label{nondicke}
\sum_{i=1}^3 \, \s{-}{i} \, P^{(D, 3)}_{3/3,3/2} \, \s{+}{i}  = |011><011| + |101><101| + |110><110| , 
\eeq
which cannot be expressed as a superposition of basis states $P^{(D, 3)}_{M, M'}$. 

If one nevertheless wants to use Dicke states to tackle such problems, one can truncate eq. (\ref{eq:damp1}) and use the operators $S_\pm$ and $S_3$ to arrive at an expression $L$ of the Lindblad form that respects the condition $Tr(L P) =0$. For $C = B/2$, i.e. neglecting the dephasing term, one obtains:
\beq \label{Dapprox}
\tilde{\Ls}^{(Z)} P = - \frac{B}{2} (1 - s) \, \left( S_+ S_- P + P S_+ S_- - 2 \, S_- P S_+ \right)  
- \frac{B}{2} s \, \left( S_- S_+ P + P S_- S_+ - 2 \, S_+ P S_- \right) 
\eeq
(Note that the natural replacement of the dephasing term by $P - S_3 P S_3$ is not possible as $S_3^2 \neq 1$ for $Z \geq 2$.) We conclude that whether or not eq. (\ref{Dapprox}) is a good approximation to eq. (\ref{eq:damp1}) (for $C=B/2$) has to be decided on a case-by-case basis by comparison with the exact solution. To find the exact solution, generalized Dicke states will prove useful.


\section{Generalized Dicke States}  \label{s:gds}

To begin with, let us study the algebraic properties of superoperators such as those occuring in eq. (\ref{eq:damp1}). We will see that these superoperators satisfy the commutation relations of the generators of the special unitary group  $SU(4)$. This observation will help us to construct appropriate basis states. These basis states are classified according to their symmetry type $Y$, which is typically represented by a Young tableau.  

\subsection{$SU(4)$ Superoperators}


We define the following superoperators:
\bea \label{eq:so} \nn
{\cal Q}_\pm  \, P := \sum_{i=1}^{Z} \, \s{\pm}{i} \, P \,  \s{\mp}{i}  \quad &,& \quad {\cal Q}_3   \, P := \4 \, \sum_{i=1}^{Z} \, \left( \s{3}{i} \,  P + P  \,  \s{3}{i}    \right)  \nn  \\ 
\Sigma_{\pm}   \, P := \sum_{i=1}^{Z} \, \s{\pm}{i}  \, P \,  \s{\pm}{i}  \quad &,& \quad \Sigma_3   \, P := \4 \, \sum_{i=1}^{Z} \, \left( \s{3}{i}  \, P - P  \,  \s{3}{i}    \right)  \nn  \\ 
{\cal M}_{\pm}   \, P := \sum_{i=1}^{Z} \, \s{\pm}{i}  \, P  \, \frac{1+ \s{3}{i}}{2}  \quad &,& \quad {\cal M}_3   \, P := \2 \, \sum_{i=1}^{Z} \, \s{3}{i} \,  P \,  \frac{1+ \s{3}{i}}{2}      \\ 
{\cal N}_{\pm}   \, P := \sum_{i=1}^{Z} \, \s{\pm}{i}  \, P  \, \frac{1- \s{3}{i}}{2}  \quad &,& \quad {\cal N}_3   \, P := \2 \, \sum_{i=1}^{Z} \, \s{3}{i}  \, P \,  \frac{1- \s{3}{i}}{2}     \nn  \\ 
{\cal U}_{\pm}   \, P := \sum_{i=1}^{Z} \, \frac{1+ \s{3}{i}}{2}  \,  P \,   \s{\mp}{i}    \quad &,& \quad {\cal U}_3   \, P := \2 \, \sum_{i=1}^{Z} \, \frac{1+ \s{3}{i}}{2}  \, P \, \s{3}{i}     \nn  \\ 
{\cal V}_{\pm}   \, P := \sum_{i=1}^{Z} \, \frac{1- \s{3}{i}}{2}   \, P  \,  \s{\mp}{i}    \quad &,& \quad {\cal V}_3   \, P := \2 \, \sum_{i=1}^{Z} \, \frac{1- \s{3}{i}}{2}  \, P \, \s{3}{i}     \nn   
\eea
$15$ of these $18$ superoperators are linearly independent. In fact, it is easy to see that ${\cal N}_3 = {\cal Q}_3  + \Sigma_3 -{\cal M}_3,$ ${\cal U}_3 = -\Sigma_3 + {\cal M}_3$ and ${\cal V}_3 = {\cal Q}_3 - {\cal M}_3$.  The $15$ independent superoperators are the generators of the Lie group $SU(4)$ (see \cite{G,S}). Table II  in Appendix A lists all commutation relations between the superoperators. Here we focus on some important commutation relations that will help us to construct the basis states. To simplify notation, let ${\bf O} :=  \{  {\cal Q}, \Sigma, {\cal M}, {\cal N}, {\cal U}, {\cal V}  \}$. We then note that for all $X \in {\bf O}$:
\bea 
\lbrack X_+ , X_- \rbrack &=& 2 \, X_3  \\ 
\lbrack X_3, X_\pm  \rbrack &=& \pm X_\pm  
\eea
That is, we have identified six $SU(2)$ subgroups. Note also that
\beq
\lbrack {\cal Q}_i , \Sigma_j \rbrack =  \lbrack {\cal M}_i , {\cal N}_j \rbrack   = \lbrack {\cal U}_i , {\cal V}_j \rbrack   = 0 \quad \forall i, j \in \{ \pm , 3 \} ,
\eeq
i.e. the $su(2)$-subalgebras for ${\cal Q}$ and ${\Sigma}$ etc. are  ``orthogonal''. Next, it is useful to study the quadratic superoperators
\beq \label{q2}
 X^2 = X_- X_+  + X_3^2 + X_3 , 
\eeq
for  all $X \in {\bf O}$. These superoperators satisfy the commutation relations 
\beq \label{xcomm}
[X^2, X'^2] =0 \quad {\rm and} \quad [X^2, X_3'] =0  \, , 
\eeq
for all $X, X' \in  {\bf O}$. Moreover, we note that
\beq 
[X_3, X_3'] = 0 \, , 
\eeq
for all $X, X' \in  {\bf O}$. Hence, there are joint eigenfunctions of ${\cal Q}^2, {\cal Q}_3,$ $\Sigma^2,$ $ \Sigma_3,$ $ {\cal M}^2,$ $ {\cal M}_3,$ $ {\cal N}^2,$ $ {\cal N}_3,$ $ {\cal U}^2,$ $ {\cal U}_3,$ $ {\cal V}^2$ and ${\cal V}_3$. Before constructing these eigenfunctions (or basis states), we note some important duality relations that hold between our superoperators. The dual conjugate $\check{L}$ to a superoperator $L$ is defined as 
\beq 
Tr \{ O (L P)  \} = Tr \{ ( \check{L} O ) P \}  
\eeq
for all states $P$ and all observables $O$ \cite{BE}. It then follows from the definition of our superoperators that
\bea
\check{\cal Q}_\pm = {\cal Q}_\mp \quad &,& \quad \check{\cal Q}_3 = {\cal Q}_3 \nn \\
\check{\Sigma}_\pm = \Sigma_\mp \quad &,& \quad \check{\Sigma}_3 = -\Sigma_3 \nn \\
\check{\cal M}_\pm = {\cal U}_\mp \quad &,& \quad \check{\cal M}_3 = {\cal U}_3  \\
\check{\cal N}_\pm = {\cal V}_\mp \quad &,& \quad \check{\cal N}_3 = {\cal V}_3 \nn .
\eea

\subsection{Basis States}

The fundamental representation of the group $SU(4)$, adapted to the present case, is explicitly given by
\bea \label{fundrep}
u := |1 ><1 | &,& d :=  | 0><0 | \nn \\
s := | 1>< 0| &,& c := |0 >< 1| .
\eea 
The names allude to the $SU(4)$ quark model. (As Georgi notes, the Flavor $SU(4)$ model was actually never very successful, but this is not relevant for the present discussion, see \cite{G}.) Table I shows how our superoperators (for $Z=1$) act on these states. All higher order representations can then be obtained from the fundamental representation and the symmetry type $Y$, which is conveniently characterized by a Young tableau comprising four rows of $p_1, p_2, p_3$ and $p_4$ boxes with $p_1+p_2+p_3+p_4 =Z$. 
\begin{center}
\begin{table}
\begin{minipage}{.25\textwidth}
\begin{tabular}{|r||c|c|c|c|} \hline
              & $u$    & $d$    & $s$  &   $c$        \\ \hline \hline
${\cal Q}_{+}$ & 0      & u      & 0    &    0  \\ \hline
${\cal Q}_{-}$ & d      & 0      & 0    &    0   \\ \hline
${\cal Q}_{3}$ & $\2 u$ & -$\2 d$ & 0    &    0  \\ \hline \hline
$\Sigma_{+}$ & 0      & 0      & 0    &    s  \\ \hline
$\Sigma_{-}$ & 0      & 0      & c    &    0   \\ \hline
$\Sigma_{3}$ & 0 & 0 & $\2 s$    &    -$\2 c$  \\ \hline
\end{tabular}
\end{minipage}
\begin{minipage}{.25\textwidth}
\begin{tabular}{|r||c|c|c|c|} \hline
              & $u$    & $d$    & $s$  &   $c$        \\ \hline \hline
${\cal M}_{+}$ & 0      & 0      & 0    &    u  \\ \hline
${\cal M}_{-}$ & c      & 0      & 0    &    0   \\ \hline
${\cal M}_{3}$ & $\2 u$ & $ 0$ & 0    &    -$\2 c$  \\ \hline \hline
${\cal N}_{+}$ & 0      & s      & 0    &    0  \\ \hline
${\cal N}_{-}$ & 0      & 0      & d    &    0   \\ \hline
${\cal N}_{3}$ & 0 & -$\2 d$ & $\2 s$    &    0  \\ \hline
\end{tabular}
\end{minipage}
\begin{minipage}{.25\textwidth}
\begin{tabular}{|r||c|c|c|c|} \hline
              & $u$    & $d$    & $s$  &   $c$        \\ \hline \hline
${\cal U}_{+}$ & 0      & 0      & u    &    0  \\ \hline
${\cal U}_{-}$ & s     & 0      & 0    &    0   \\ \hline
${\cal U}_{3}$ & $\2 u$ & 0 & -$\2 s$    &    0  \\ \hline \hline
${\cal V}_{+}$ & 0      & c      & 0    &    0  \\ \hline
${\cal V}_{-}$ & 0      & 0      & 0    &    d   \\ \hline
${\cal V}_{3}$ & 0 & -$\2 d$ & 0    &    $ \2 c$  \\ \hline
\end{tabular}
\end{minipage}
\caption{The action of the $SU(4)$ superoperators on the elements of the fundamental representation.}
\end{table}
\end{center}

It is important to note that the symmetry type $Y$ of a state is preserved under the action of the superoperators defined in  (\ref{eq:so}): if one of the superoperators  (or a linear combination thereof) is applied to a state of symmetry type $Y$, then the resulting state also has the symmetry type $Y$. Hence, the non-unitary dynamics preserves the symmetry type $Y$ as the terms on the RHS of eq.  (\ref{eq:damp1}) can be expressed in terms of the superoperators defined in (\ref{eq:so}). The same holds for the unitary dynamics. To see this we note that the atomic part of the Hamiltonian $H$ involves terms proportional to $S_\pm$ and $S_3$. The commutator $\left[ H, P \right]$, which captures the unitary dynamics in the quantum master equation, then involves terms proportional to $S_\pm \, P = ({\cal M}_\pm + {\cal N}_\pm) \, P$, $P \, S_\pm =  ({\cal U}_\pm + {\cal V}_\pm) \, P$, $S_3 \, P = ({\cal Q}_3 + \Sigma_3) \, P$ and $P \, S_3 =  ({\cal Q}_3 - \Sigma_3) \, P$.

Let us now identify the basis states. For the configuration $u^\alpha \, d^\beta \, s^\gamma \, c^\delta$ (with $\alpha + \beta + \gamma + \delta = Z$), these are given by
\beq \label{state}
P_Y= {\cal S}^Y \, (u^\alpha \, d^\beta \, s^\gamma \, c^\delta), 
\eeq
where the symmetrizer ${\cal S}^Y$ makes sure that the state has symmetry type $Y.$  ${\cal S}^Y$ is the product of symmetrizers and anti-symmetrizers which can be read-off from the corresponding Young tableau. (Note that the symmetrizers have to be applied first, then the antisymmetrizers, so that the resulting state is antisymmetrical in the corresponding indices, but not necessarily symmetrical in all indices which were symmetrized  \cite{S}.) 

To construct the basis states, one can also start with a state of highest weight (for a given symmetry type $Y$) and apply the raising and lowering operators ${\cal Q}_\pm$ etc. to generate all other states.  For full symmetry (i.e. for the Young tableau $(Z, 0, 0, 0)$), the states of highest weight are $u^Z, d^Z, s^Z$ and $c^Z$. 

We note that only fully symmetrical states which exclusively involve $u$'s and $d$'s have a non-vanishing trace. This follows from two considerations. First, the explicit form of the fundamental representation (see eq. (\ref{fundrep})) implies that 
\beq \label{tracefund}
Tr(u) = Tr(d) =1 \quad {\rm and} \quad Tr(s) = Tr(c) =0.
\eeq
Hence, only composite states containing $u$'s and $d$'s have a non-vanishing trace. Second, because of the previous consideration, Young tableaux for non-fully-symmetrical states have exactly two rows. (A third or forth row would require elementary $s$ or $c$ states to fulfill the symmetry requirement.) Hence, each term in the corresponding state contains at least one antisymmetrized $u/d$ pair, and so the trace of such states vanishes. 

In the remainder of this paper, we focus on fully symmetrical states and leave a detailed analysis of states of mixed symmetry and fully antisymmetrical states to a sequel to this paper. We reserve the name {\em generalized Dicke states} for fully symmetrical basis states of trace 1. (For a different way to generalize Dicke states, see \cite{PR}.) Strictly speaking, we should not use the word ``state'' at all if the trace vanishes. For convenience, however, we will do so anyway. This also makes sense as states of vanishing trace can be created by the dynamics as admixtures to states of trace 1.

\subsection{Fully Symmetrical States}


For ease of notation, we denote the symmetrizer for fully symmetrical states by ${\cal S}$, so that the basis states are given by
\beq \label{states}
P^{(s)} = {\cal S} \, (u^\alpha \, d^\beta \, s^\gamma \, c^\delta) .
\eeq
To specify suitable quantum numbers that label the basis states, we note that ${\cal S}$ commutes with all our superoperators. Using  Table I, we then derive:
\bea \label{qpm}
{\cal Q}_- {\cal Q}_+ \, P^{(s)}  = (\alpha +1) \, \beta \, P^{(s)} \quad &,& \quad {\cal Q}_3 \, P^{(s)} = \2 \, (\alpha - \beta) \, P^{(s)} \nn \\
\Sigma_- \Sigma_+ \, P^{(s)}  = (\gamma +1) \, \delta \, P^{(s)} \quad &,& \quad \Sigma_3 \, P^{(s)} = \2 \, (\gamma - \delta) \, P^{(s)} \nn \\
{\cal M}_- {\cal M}_+ \, P^{(s)}  = (\alpha +1) \, \delta \, P^{(s)} \quad &,& \quad {\cal M}_3 \, P^{(s)} = \2 \, (\alpha - \delta) \, P^{(s)}  \\
{\cal N}_- {\cal N}_+ \, P^{(s)}  = \beta \, (\gamma +1) \,  P^{(s)} \quad &,& \quad {\cal N}_3 \, P^{(s)} = \2 \, (-\beta + \gamma) \, P^{(s)} \nn \\
{\cal U}_- {\cal U}_+ \, P^{(s)}  = (\alpha +1) \, \gamma \, P^{(s)} \quad &,& \quad {\cal U}_3 \, P^{(s)} = \2 \, (\alpha - \gamma) \, P^{(s)} \nn \\
{\cal V}_- {\cal V}_+ \, P^{(s)}  = \beta \,  (\delta +1) \, P^{(s)} \quad &,& \quad {\cal V}_3 \, P^{(s)} = \2 \, (-\beta + \delta) \, P^{(s)} \nn 
\eea
With eq. (\ref{q2}) and requesting (or defining), as usual, that $X^2 \, P^{(s)} = x \, (x+1) \, P^{(s)}$ and $X_3 \, P^{(s)} = x_3 \, P^{(s)}$ for all $X \in {\bf O}$, we obtain from the first line of eqs. (\ref{qpm}):
\beq \label{alphabeta}
(\alpha +1) \, \beta + \4 \, (\alpha - \beta)^2 + \2 \, (\alpha - \beta) =: q(q+1) \, .
\eeq
Hence (neglecting the negative solution)
%
\beq \label{qeq}
q = \2 \, (\alpha + \beta)  \, .
\eeq
Moreover, 
\beq \label{q3}
q_3 = \2 \, (\alpha - \beta)  \, .
\eeq
Similarly, we obtain:
\bea \label{eigenmnuv}
\sigma =  \2 \, (\gamma + \delta)  \quad &,& \quad \sigma_3 = \2 \, (\gamma - \delta) \nn \\
m =  \2 \, (\alpha + \delta)  \quad &,& \quad m_3 = \2 \, (\alpha - \delta) \nn \\
n =  \2 \, (\beta + \gamma)  \quad &,& \quad n_3 = \2 \, (-\beta + \gamma)  \\
u =  \2 \, (\alpha + \gamma)  \quad &,& \quad u_3 = \2 \, (\alpha - \gamma) \nn \\
v =  \2 \, (\beta + \delta)  \quad &,& \quad v_3 = \2 \, (-\beta + \delta) \nn 
\eea 
Note that
\beq \label{qsigsym}
q+ \sigma = m + n = u + v = Z/2 \, .
\eeq
From this we conclude that each term in a fully symmetrical basis state has  (i) the same numbers of 1's ($=2m$) and 0's ($=2n$) on the RHS (i.e. the $< \cdot \vert $'s) of the density operator, (ii) the same numbers of 1's ($=2u$) and 0's ($=2v$) on the LHS (i.e. the $\vert \cdot >$'s) of the density operator, and (iii) the same number of  $0-0$ or $1-1$ coincidences ($=2q$) and  $1-0$ or $0-1$ non-coincidences ($= 2\sigma$) on the LHS and the RHS of each term in a basis state. These observations give our quantum numbers some intuitive meaning.


All fully symmetrical basis states for a given number $Z$ of atoms can then be characterized by three quantum numbers. We choose $q, q_3$ and $\sigma_3$ from which we can calculate $\alpha, \beta, \gamma$ and $\delta$ in turn:
\bea \label{abcd}
\alpha = q + q_3 \quad &,& \quad  \beta = q - q_3 \nn \\
 \gamma = \sigma + \sigma_3 \quad &,& \quad \delta = \sigma - \sigma_3 ,
\eea
with $\sigma = Z/2 - q$. Let us now explore the range of the values of these quantum numbers. From $\lbrack {\cal Q}_3, {\cal Q}_\pm  \rbrack = \pm {\cal Q}_\pm$ we conclude that ${\cal Q}_+$ raises $q_3$ by one unit until $\beta =0$, i.e. until $q_3 = q$. Similarly, ${\cal Q}_-$ lowers $q_3$ by one unit until $\alpha=0$, i.e. until $q_3 = -q$. Hence, $q_3$ ranges from $-q$ to $q$. Similarly, $\sigma_3$ ranges from $-\sigma$ to $\sigma$. From eq. (\ref{qeq}) we conclude that $q$ ranges from $0$ to $Z/2$ in steps of $1/2$ unit. We denote the basis states by $P^{(Z)}_{q, q_3,\si_3}$ and arrive at the following list of the three quantum numbers that characterize them:
\begin{enumerate}
\item $q: {\cal Q}^2 \, P^{(Z)}_{q, q_3,\si_3} = q(q+1) \, P^{(Z)}_{q, q_3,\si_3}$, with $q= 0, 1/2, \dots , Z/2$
\item $q_3: {\cal Q}_3  \, P^{(Z)}_{q, q_3,\si_3} = q_3 \, P^{(Z)}_{q, q_3,\si_3}$,  with $-q \le q_3 \le q$
\item $\si_3: \Sigma_3 \, P^{(Z)}_{q, q_3,\si_3} = \sigma_3 \, P^{(Z)}_{q, q_3,\si_3}$, with $- \sigma \le \sigma_3 \le \sigma$ and $\sigma = Z/2 - q$
\end{enumerate}

Note that the quantum numbers $m, m_3$ etc. can be expressed in terms of $q, q_3$ and $\sigma_3$. Given the symmetry of our approach, we could have also used the eigenvalues of the ${\cal M}/{\cal N}$ or ${\cal U}/{\cal V}$ superoperators to label our basis states. The ${\cal Q}/\Sigma$ labeling is, however, most convenient  for the applications we have in mind.

Generalized Dicke states are fully symmetrical basis states of trace 1. For them, $\gamma = \delta = 0$ and hence $\alpha + \beta = Z$. Using eq. (\ref{qeq}), we conclude that generalized Dicke states are characterized by the quantum numbers $q = Z/2, q_3=-q,\dots, q$ and $\sigma_3 = 0$. All other fully symmetrical basis states have a vanishing trace.  To see how the basis states look like, we calculate a generalized Dicke state with highest weight,
\beq
P^{(3)}_{3/2, 3/2,0} =  {\cal S} \, (u^3) =u^3 =  |111>  <1 1 1| \, ,
\eeq
and the state
\bea 
P^{(3)}_{1,1,1/2} 
&=&  {\cal S} \, (u^2 s) = \frac{1}{3} \, (u^2 s + u s u + s u^2) \nn \\
&=& \frac{1}{3} \, |111> \left( <1 1 0|  + <1 0 1|  + <0 1 1|  \right) \nn .
\eea

Applying ${\cal Q}_-$ to $P^{(3)}_{3/2, 3/2,0}$, we obtain $|011><011| + |101><101| + |110><110|$ (see eq. (\ref{nondicke})), which equals $3 \, P^{(3)}_{3/2, 1/2,0}$. Let us now study the action of the raising and lowering operators on our basis states more generally. For ${\cal Q}_\pm$ and $\Sigma_\pm$, we obtain: 
\bea
{\cal Q}_{\pm} \, P_{q, q_3, \sigma_3}^{(Z)} &=& (q \mp q_3) \, P_{q, q_3 \pm 1 , \sigma_3}^{(Z)} \nn \\
\Sigma_{\pm} \, P_{q, q_3, \sigma_3}^{(Z)} &=& (\sigma \mp \sigma_3) \, P_{q, q_3, \sigma_3 \pm 1}^{(Z)} 
\eea
To calculate the action of ${\cal M}_+$ on our basis states, we observe that
\beq
{\cal M}_+ \,  {\cal S} \, (u^\alpha \, d^\beta \, s^\gamma \, c^\delta) = \delta \,  {\cal S} \, (u^{\alpha +1} \, d^\beta \, s^\gamma \, c^{\delta -1}) =: \delta \,  {\cal S} \, (u^{\alpha '} \, d^{\beta '} \, s^{\gamma '} \, c^{\delta '}) .
\eeq
Hence, with eqs. (\ref{abcd}), $q' = q + 1/2, q_3' = q_3 + 1/2$ and $\sigma_3' = \sigma_3 + 1/2.$ Similarly for the other superoperators, so that we finally obtain:
\bea
{\cal M}_{\pm} \, P_{q, q_3, \sigma_3}^{(Z)} &=& (m \mp m_3) \, P_{q \pm 1/2, q_3 \pm 1/2, \sigma_3  \pm 1/2}^{(Z)} \nn \\
{\cal N}_{\pm} \, P_{q, q_3, \sigma_3}^{(Z)} &=& (n \mp n_3) \, P_{q  \mp 1/2, q_3  \pm 1/2, \sigma_3  \pm 1/2}^{(Z)}  \\
{\cal U}_{\pm} \, P_{q, q_3, \sigma_3}^{(Z)} &=& (u \mp u_3) \, P_{q \mp 1/2, q_3  \pm 1/2, \sigma_3  \mp 1/2}^{(Z)} \nn \\
{\cal V}_{\pm} \, P_{q, q_3, \sigma_3}^{(Z)} &=& (v \mp v_3) \, P_{q \pm 1/2, q_3  \pm 1/2, \sigma_3  \mp 1/2}^{(Z)} \nn 
\eea
The expressions for the new values of $q_3$ and $\sigma_3$ after a certain raising or lowering operator acted on the corresponding state can also be obtained from commutation relations such as $\lbrack {\cal Q}_3, {\cal M}_\pm  \rbrack = \pm 1/2 \, {\cal M}_\pm$ and $\lbrack \Sigma_3, {\cal M}_\pm  \rbrack = \pm 1/2 \, {\cal M}_\pm$. However, calculating in this way how the value of $q$ changes is more difficult  as  ${\cal Q}^2$ is a quadratic operator. But a trick helps. Inspired by eq. (\ref{qeq}), we define the superoperator $\tilde{\cal Q}$ by
\beq
\tilde{\cal Q} := \4 \, \left( Z + {\cal Q}_{33}  \right)
\eeq
with 
\beq
{\cal Q}_{33} \, P := \sum_{i=1}^{Z} \, \s{3}{i} \, P \,  \s{3}{i} = \left( 4 \, {\cal M}_{3} - 2 \, \left( {\cal Q}_{3}  + \Sigma_3  \right) \right) P.
\eeq

It is easy to see that 
\beq \label{comqtilde}
[\tilde{\cal Q}, {\cal Q}_{i}] = 0 \quad {\rm for} \ i \in \{ \pm, 3\}
\eeq
and that, for fully symmetrical states, 
\beq \label{qtildeeigen}
\tilde{\cal Q} \, P_{q, q_3, \sigma_3}^{(Z)} = q \, P_{q, q_3, \sigma_3}^{(Z)}. 
\eeq
Besides, $\lbrack \tilde{\cal Q}, {\cal M}_\pm  \rbrack = \pm 1/2 \, {\cal M}_\pm$. Hence, ${\cal M}_{\pm}$ raises/lowers the value of the quantum number $q$ by $1/2$ unit for fully symmetrical states. Similarly, linearized superoperators $\tilde{X}$ (for all $X \in {\bf O}$) can be constructed in a straight forward way from eqs. (\ref{eigenmnuv}). Using these superoperators simplifies calculations considerably, but note that they only work for fully symmetrical states.  

Next, we define the dual states,
\beq
\PPD = P_{q, q_3 , - \sigma_3}^{(Z)} ,
\eeq

which satisfy the biorthogonality relation
\beq
Tr(\PPD \, P_{q', q'_3 , \sigma'_3}^{(Z)})
=  \delta_{q \, q'} \, \delta_{q_3 \, q'_3} \, \delta_{\si_3 \, \si'_3} .
\eeq

To conclude this section, we determine the dimension of the space containing all fully symmetrical states. By simple counting or using the {\em hook rule} for Young tableaux, we obtain
\beq \label{dimsym}
D_{GD}(Z) = \frac{1}{6} \,  (Z+1)(Z+2)(Z+3) .
\eeq

That is, $D_{GD}(5) = 56, D_{GD}(10) = 286$ and $D_{GD}(20) = 1771$. These numbers have to be compared to $4^Z$ if one proceeds by brute force. Note, for example, that  $4^{10}  \approx 3670 \times 286$, i.e. for ten atoms, we save already a factor of 3670. Compared to the number of Dicke states for the same number of atoms (see eq. ({\ref{dimdick})), we obtain, for example, $D_{GD}(10)/D_{D}(10) \approx 2.4$ and $D_{GD}(50)/D_{D}(50) \approx 9$, which indicates only a relatively moderate increase. Hence, our basis states will be useful for numerical solutions of quantum master equations. In the next section we show that some quantum master equations can even be solved analytically.  

\section{Solving Quantum Master Equations}  \label{s:solving}

Consider first a quantum master equation with a Lindblad operator given by eq. (\ref{eq:damp1}) for $C=B/2$ (i.e. without the dephasing term). That is, we want to solve the equation
\bea \label{eq:master}
\frac{d P}{d t}  =   \Ls^{(Z)} P
&=& - \frac{B}{2} (1 - s) \  \sum_{i = 1}^{Z}  \,
[\s{+}{i} \s{-}{i} P + P \, \s{+}{i} \s{-}{i} - 2 \s{-}{i} P \s{+}{i}]  \nn  \\
&-& \frac{B}{2} s \ \sum_{i = 1}^{Z} \
[\s{-}{i} \s{+}{i} P + P \, \s{-}{i} \s{+}{i} - 2 \s{+}{i} P \s{-}{i}]  .
\eea

Using the superoperators defined in (\ref{eq:so}),  this equation can be compactly written as
\beq \label{masterq}
\frac{1}{B} \, \frac{d P}{d t}  =  \left[ - Z/2  + (1-s) \ {\cal Q}_-   - (1-2s) \ {\cal Q}_3 +  s \ {\cal Q}_+  \right] \, P .
\eeq
We introduce the dimensionless time $\tau := B \, t$ and solve eq. (\ref{masterq}) formally: 
\beq
P(\tau) = e^{- Z/2 \, \tau} \, e^{ \left(  (1-s) \ {\cal Q}_-   - (1-2s) \ {\cal Q}_3 +  s \ {\cal Q}_+  \right) \, \tau} \, P(0)
\eeq
Here $P(0)$ is the initial state of the system. We now use the Baker-Campbell-Hausdorff formula to factorize the second exponential and obtain the following two formulas (which one is most convenient will depend on the initial state).
\bea \label{solution1} 
P(\tau) &=& e^{- Z/2 \, \tau} \, e^{ A_s(\tau) {\cal Q}_+} \, e^{B_s(\tau) {\cal Q}_3} \, e^{C_s(\tau) {\cal Q}_-} \, P(0)  \label{BCH1} \\
&=& e^{- Z/2 \, \tau} \, e^{ D_s(\tau) {\cal Q}_-} \, e^{E_s(\tau) {\cal Q}_3} \, e^{F_s(\tau) {\cal Q}_+}  \, P(0) ,  \label{BCH2} \label{solution2}
\eea
with
\bea
A_s(\tau) &=&  \frac{ s \, f({\tau})}{1- s \, f({\tau})} \nn  \\
B_s(\tau) &=& - \tau - 2 \log (1-s \, f({\tau}) ) \nn  \\
C_s(\tau) &=& \frac{(1-s) \, f({\tau})}{1- s\, f({\tau})} \nn  \\
D_s(\tau) &=&   \frac{(1-s) \, f({\tau})}{1-(1-s)\, f({\tau})}  = A_{1-s}(\tau)  \\
E_s(\tau) &=&  \tau + 2 \log (1-(1-s)\, f({\tau})) = -B_{1-s}(\tau) \nn  \\
F_s(\tau) &=&   \frac{s \, f({\tau})}{1-(1-s)\, f({\tau})} = C_{1-s}(\tau) \nn
\eea
and
\beq
f(\tau) := 1- e^{-\tau}. 
\eeq

Master equations involving other or more superoperators from (\ref{eq:so}) can be formally solved in a similar way. Here we only show how the dephasing term from eq. (\ref{eq:damp1}) can be included. To do so, we first note that
\bea \label{eq:damp}
\Ls^{(Z, deph)} P &:=& - \frac{2 C - B}{4} \  \sum_{i = 1}^{Z} \ [P - \s{3}{i} P \  \s{3}{i}]  \nn \\
&=& B \, (1- 2 \tilde{C}) \, (Z/2 - \tilde{\cal Q}) \, P \nn \\
&=:& B\, (1- 2 \tilde{C}) \, \tilde{\Sigma} \, P ,
\eea
with $\tilde{C} := C/B$. Next, we recall that $\tilde{\cal Q}$ (and hence $\tilde{\Sigma}$) commutes with ${\cal Q}_\pm$ and ${\cal Q}_3$ (see eq. (\ref{comqtilde})). Hence, the solution of the quantum master equation including the dephasing term simply adds the factor 
\beq \label{pdeph}
P_{deph} (\tau)= e^{(1-2 \tilde{C}) (Z/2 - \tilde{\cal Q}) \, \tau}
\eeq
to the right hand sides of eqs. (\ref{solution1}) and (\ref{solution2}). Note that generalized Dicke states (for which $q=Z/2$) are not affected by the dephasing term. In the following illustration and the applications in the next section, we disregard the dephasing term.

To illustrate the method developed in this section, let us calculate the time evolution of the fully symmetrical state $P_{q, q_3, \sigma_3}^{(Z)}$ under the influence of pure decoherence, i.e. for $s=0$. Using eq. (\ref{solution1}), the resulting state is given by 
\beq
P (\tau) = e^{-Z / 2 \, \tau }  \,  e^{- \tau {\cal Q}_{3} } \ e^{f(\tau) \cQ_{-}  }  \ P_{q, q_3, \sigma_3}^{(Z)} \, ,
\eeq
and after some algebra, we obtain the compact expression
\beq \label{decay}
P (\tau) = e^{-Z / 2 \, \tau } \, \sum_{k=0}^{q+q_3} \, \binomial{q+q_3}{k} \,  f^k(\tau) \,  (1-f(\tau))^{q_3-k}  \, P_{q, q_3-k, \sigma_3}^{(Z)} \, .
\eeq

Using eq. (\ref{decay}), we can easily calculate the time evolution of the atomic inversion $<S_3> = <{\cal Q}_3> = Tr({\cal Q}_3 \, P (\tau))$ of a decaying fully symmetrical state. For the initial state $P_{Z/2, 0, 0}^{(Z)}$ (with even $Z$), we obtain $<S_3>/Z = 1/2 \, (e^{-\tau} -1)$. For the initial state $P_{Z/2, Z/2, 0}^{(Z)} = |1\dots 1><1\dots 1|$, we obtain $<S_3>/Z = e^{-\tau} -1/2$. It is instructive to compare the latter result  with the expression that one obtains if one uses Dicke states and replaces the present Lindblad operator by the one given in eq.  (\ref{Dapprox}) for $s=0$. In the Dicke basis, the initial state is given by $P^{(D, Z)}_{Z/2, Z/2}$ and the resulting atomic inversion for $Z=2$ is $<S_3>/2 = (1+ \tau) \, e^{-2 \tau} -1/2$, which suggests a much faster decay.

\section{Applications}  \label{s:apps}

Let us now consider three applications of our formalism. We first study the propagation of a Bell state for an arbitrary pumping parameter $s$ (Sec. \ref{s:bell}). Then we calculate the decay of a GHZ state (Sec. \ref{s:GHZ}). Finally, we calculate the eigenvalues of the Lindblad operator (Sec. \ref{s:evs}).

\subsection{The Propagation of a Bell State} \label{s:bell}

We calculate the propagation of the Bell state $\vert \Psi^+ > := \frac{1}{\sqrt{2}} \, \left( \vert 01 > + \vert 10 > \right)$ for arbitrary $s$. To do so, we first note that $P_B(0):= \vert \Psi^+ > < \Psi^+ \vert = {\cal S} (u d + sc) = P^{(2)}_{1,0,0} + P^{(2)}_{0,0,0}$ and obtain from eq. (\ref{decay}) for the time evolution of $P_B(0)$:
\beq \label{bellt}
P_B(\tau) = b_1(\tau) \, P^{(2)}_{1,1,0} + b_2(\tau) \, P^{(2)}_{1,0,0} + b_3(\tau) \, P^{(2)}_{1,-1,0} + b_4(\tau) \, P^{(2)}_{0,0,0}
\eeq
The weights are given by $b_1(\tau) = s \, f({\tau})\, [1-(1-s) \, f({\tau})]$, $b_2(\tau) =  1- f({\tau})\, [1-2 s (1-s) \, f({\tau})]$, $b_3(\tau) = (1-s) \, f({\tau})\,  [1-s \, f({\tau})]$ and $b_4(\tau) = 1-f({\tau})$. For $s=1/2$, $P_B(\tau)$ simplifies to
\beq
P_B(\tau)  = \frac{1}{4} \, (1-e^{-2 \tau} ) \, P^{(2)}_{1,1,0} + \frac{1}{2} \, (1+ e^{-2 \tau}) \, P^{(2)}_{1,0,0} + \frac{1}{4} \, (1-e^{-2\tau}) \, P^{(2)}_{1,-1,0} + e^{-\tau} \, P^{(2)}_{0,0,0},
\eeq
which converges asymptotically  to $1/4  \, P^{(2)}_{1,1,0} + 1/2 \, P^{(2)}_{1,0,0}  + 1/4 \, P^{(2)}_{1,-1,0}$. More generally, $P_B(\tau)$ converges asymptotically to
\beq
P_B(\tau)  = s^2 \, P^{(2)}_{1,1,0} + 2 s (1-s) \, P^{(2)}_{1,0,0}  +  (1-s)^2 \, P^{(2)}_{1,-1,0}.
\eeq

Let us now calculate the von Neumann entropy of $P_{B} (\tau)$ as a function of time $\tau$ for different values of the pumping parameter $s$. To do so, we diagonalize the density matrix from eq.  (\ref{bellt}) and calculate its eigenvalues $\lambda_i$ for $i=1,\dots, 4$. The von Neumann entropy is then given by $S_B (\tau)= - Tr(P_{B} (\tau) \, \log P_{B} (\tau)) = - \sum_{i=1}^4 \lambda_i \, \log \lambda_i$. (The base of the log is 2.) It is depicted in Figure 1.  


\begin{figure}[h]
\begin{center}
\includegraphics[scale=1.0]{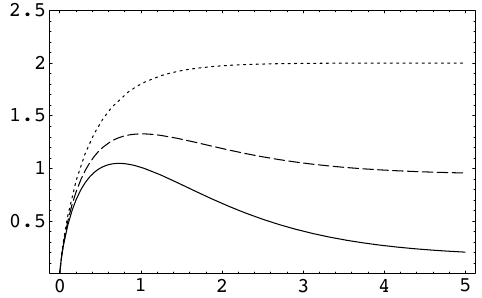}\label{f2} 
\caption{The von Neumann entropy $S_B (\tau)$ of $P_{B} (\tau)$ as a function of  time $\tau$ for $s=.5$ (dotted), $s=.1$ (dashed) and $s=.9$ (solid).}
\end{center}
\end{figure}

\subsection{The Decay of a GHZ State} \label{s:GHZ}

Next, we consider the case of pure damping (i.e. $s=0$), and propagate a GHZ state through time. To do so, we express the density operator corresponding to the GHZ state
\beq
\vert {\rm GHZ} > = \frac{1}{\sqrt{2}} \, \left(  \vert  1 1 1> -  \vert 0 0 0 > \right) 
\eeq
 in our fully symmetrical basis for $Z = 3$.  We obtain:
\bea
P_{GHZ} &=& \vert {\rm GHZ > < GHZ} \vert  \nn \\
&=&  1/2 \left( \vert  1 1 1> < 1 1 1    \vert + \vert 0 0 0 > < 0 0 0 \vert  -  \vert 1 1 1 > <  0 0 0   \vert  - \vert 0 0 0 > < 1 1 1    \vert  \right)  \nn  \\
&=& 1/2 \left(u^3 + d^3 -s^3 - c^3 \right) \nn \\
& =& 1/2 \left(P^{(3)}_{3/2, 3/2, 0} + P^{(3)}_{3/2, -3/2,0} - P^{(3)}_{0,0,3/2}  - P^{(3)}_{0,0,-3/2} \right) \nn
\eea
For $s=0$, we then have to calculate (see eq. (\ref{decay}))
\bdm
P_{GHZ}  (\tau) = e^{-3 / 2 \, \tau }  \,  e^{- \tau {\cal Q}_{3} } \ e^{f(\tau) \cQ_{-}  }  \ P_{GHZ} 
\edm
and obtain:
\bea \label{GHZ}
P_{GHZ} (\tau) =  c_1(\tau) \, P^{(3)}_{3/2, 3/2, 0} + c_2(\tau) \, P^{(3)}_{3/2,  1/2,0} + c_3(\tau) \, P^{(3)}_{3/2,  -1/2,0} \nn \\ 
+ c_4(\tau) \, P^{(3)}_{3/2,  -3/2,0} - c_5(\tau) \, \left( P^{(3)}_{0,0,3/2} +  P^{(3)}_{0,0,-3/2} \right)  ,  
\eea
with the weights $c_1(\tau) =   1/2 \, e^{-3 \tau}, c_2(\tau) =   3/2 \, e^{-2 \tau} \, f(\tau), c_3(\tau) =   3/2 \, e^{- \tau} \,  f(\tau)^2,  c_4(\tau) =    1/2 \, (1+  f(\tau)^3)$ and $c_5(\tau) = 1/2  \, e^{-3/2 \tau}$.  The corresponding von Neumann entropy $S_{GHZ} (\tau)$ is depicted in Figure 2. It vanishes, once the system reaches its asymptotic state with all atoms in the ground state.

\begin{figure}[h]
\begin{center}
\includegraphics[scale=1.0]{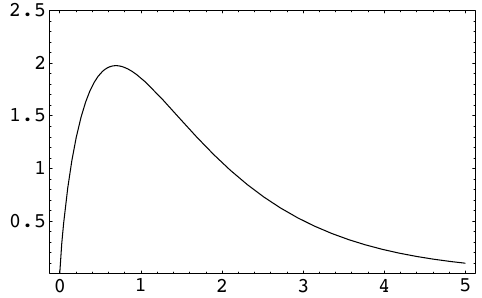}\label{f3} 
\caption{The von Neumann entropy $S_{GHZ} (\tau)$ of $P_{GHZ} (\tau)$ as a function of  time $\tau$.}
\end{center}
\end{figure}

\subsection{Eigenvalues of the Lindblad Operator} \label{s:evs}

In this subsection, we calculate the eigenvalues of the Lindblad operator (again with $C=B/2$). That is, we  solve the equation
\beq \label{eigendamp}
L \, P = B \left[ - Z/2 +  (1-s) \ {\cal Q}_- - (1-2s) \ {\cal Q}_3  +  s \ {\cal Q}_+ \right]  P = \lambda \, P .
\eeq
As only states with $q=Z/2$ have a non-vanishing trace (and as the Lindblad operator does not change $q$), we expand $P$ in terms of states with $q=Z/2$:
\beq \label{ansatz}
P = \sum_{q_3 = -Z/2}^{Z/2} \, c^{(Z)}_{q_3} \, P_{ Z/2, q_3 , 0}^{(Z)}
\eeq

If we insert eq. (\ref{ansatz}) in eq. (\ref{eigendamp}), then we obtain a recursion equation for the expansion coefficients $c^{(Z)}_{q_3}$. This equation can be brought into matrix form, which leaves us with the calculation of the eigenvalues $\lambda$ and the corresponding eigenvectors. We obtain
\beq
\lambda/B = 0, -1, -2, \dots, -Z
\eeq
and the corresponding eigenvectors using the software {\em Mathematica}. If we insert the eigenvectors for $\lambda/B = -1, -2, \dots, -Z$ into eq. (\ref{ansatz}), then we see that the corresponding states have trace zero. However, the states corresponding to the eigenvectors for $\lambda/B = 0$ have a finite trace and are therefore of direct physical interest. In this case, the normalized coefficients are given by the binomial distribution
\beq
c^{(Z)}_{q_3} = \binomial{Z}{Z/2 + q_3} \, s^{Z/2 + q_3} \, (1-s)^{Z/2 - q_3} \, .
\eeq
For $s=0$, the corresponding state is simply $P_{ Z/2, -Z/2, 0}^{(Z)} = d^Z$ and for $s=1$, the corresponding state is $P_{ Z/2, Z/2, 0}^{(Z)} = u^Z$. For $s=1/2$ the corresponding state can be compactly written as $\left[ (u+d)/2  \right]^Z$.

\section{Conclusions}  \label{s:conclusions}

We have presented a group-theoretical superoperator method to solve quantum master equations for a finite number of atoms (or quits). The method can be applied to many further problems in quantum optics and quantum information theory. Studies that used the method proposed in this paper include \cite{App1,App2,App3,App4,App5,App6,App7}. Here are a few more examples. For quantum optical applications, it is interesting to couple the atoms to a radiation field (using the Tavis-Cummings model) and to study, for example, the physics of few-atom lasers as well as phenomena such as resonance flourescence and optical bistability in $Z$-atom systems in a cavity. The resulting equations have to be solved numerically, but  the complexity of the problems will be considerably reduced with the help of the basis states proposed in this paper. On the more theoretical side and with an eye on applications in quantum information theory, it will be interesting to construct $SU(4)$ coherent states and study their decoherence times. Moreover, one can ask which many-atom states are especially stable under the influence of decoherence, and which not? We believe that our method will be a valuable tool in these studies. While this paper focused on fully symmetrical states, states of mixed symmetry and fully antisymmetrical states can also be prepared experimentally (as admixtures to generalized Dicke states). In a sequel to this paper, we will develop the theory of these states and study some of its applications.




\appendix

\section{Proofs and commutators}

 In Table II, we list the commutation relations of the superoperators ${\cal Q}_\pm$ etc. For symmetry reasons, we list the commutation relations of all 18 superoperators, although only 15 of them are linearly independent. For reasons of space, we use the abbreviation $\xi = 1/2$.

To calculate the commutators $[X^2, X'^2] =0$ for all $X, X' \in  {\bf O}$, we first conclude from the commutation relations from Table II that 
\bdm
[X^2, X'^2] = [X_- X_+, X'_- X'_+] \, .
\edm
To calculate the remaining commutator, we first consider the case $X = {\cal Q}$ and $X' = {\cal M}$ and obtain:
\bdm
[{\cal Q}_- {\cal Q}_+, {\cal M}_- {\cal M}_+] =  {\cal M}_- {\cal V}_-{\cal Q}_+ - {\cal Q}_- {\cal V}_+{\cal M}_+  
\edm
This expression cannot be simplified further by using commutation relations. Using the explicit definitions of the superoperators from eq. (\ref{eq:so}), we finally obtain
\bdm
[{\cal Q}^2, {\cal M}^2] = 0 \, .
\edm
Similarly for the other superoperators.

\begin{table}
\begin{center}
\begin{tabular}{|r||c|c|c|c|c|c||c|c|c|c|c|c||c|c|c|c|c|c|} \hline
              & ${\cal Q}_{+}$    & ${\cal Q}_{-}$ & ${\cal Q}_{3}$  &   $\Sigma_{+}$   &   $\Sigma_{-}$  &   $\Sigma_{3}$  & ${\cal M}_{+}$    & ${\cal M}_{-}$    & ${\cal M}_{3}$  & ${\cal N}_{+}$ & ${\cal N}_{-}$    & ${\cal N}_{3}$  & ${\cal U}_{+}$  & ${\cal U}_{-}$  & ${\cal U}_{3}$    & ${\cal V}_{+}$    & ${\cal V}_{-}$ & ${\cal V}_{3}$  \\ \hline \hline
${\cal Q}_{+}$ & 0      & $2 {\cal Q}_{3}$  & -${\cal Q}_{+}$ & 0 & 0    &  0&  0 & -${\cal V}_{+}$  & -$\xi {\cal Q}_{+}$  & 0 &  ${\cal U}_{+}$ & -$\xi {\cal Q}_{+}$ & 0 & -${\cal N}_{+}$ & -$\xi {\cal Q}_{+}$ & 0    &   ${\cal M}_{+}$ & -$\xi {\cal Q}_{+}$     \\ \hline
${\cal Q}_{-}$ &   & 0 & ${\cal Q}_{-}$ & 0  &  0 & 0 & ${\cal V}_{-}$ & 0 & $\xi {\cal Q}_{-}$ & -${\cal U}_{-}$ & 0 & $\xi {\cal Q}_{-}$ & ${\cal N}_{-}$ & 0 & $\xi {\cal Q}_{-}$ & -${\cal M}_{-}$ &    0 & $\xi {\cal Q}_{-}$    \\ \hline
${\cal Q}_{3}$ &  &  & 0 & 0& 0 & 0 & $\xi {\cal M}_{+}$ & -$\xi {\cal M}_{-}$ & 0 & $\xi {\cal N}_{+}$ & -$\xi {\cal N}_{-}$    &    0& $\xi {\cal U}_{+}$ & -$\xi {\cal U}_{-}$ & 0 &    $\xi {\cal V}_{+}$ & -$\xi {\cal V}_{-}$    &    0 \\ \hline 
$\Sigma_{+}$ &       &       &     &    0 & $2 \Sigma_{3}$    &    -$\Sigma_{+}$ & 0      &   ${\cal U}_{-}$    & -$\xi \Sigma_+$    &    0 & -${\cal V}_{-}$    & -$\xi \Sigma_+$ & -${\cal M}_{+}$      & 0      & $\xi \Sigma_+$    &    ${\cal N}_{+}$ & 0    &    $\xi \Sigma_+$\\ \hline
$\Sigma_{-}$ &      &       &   &  &    0  & $\Sigma_-$    &    ${\cal U}_{+}$ & 0      & $\xi \Sigma_-$      & ${\cal V}_{+}$    &    0 & $\xi \Sigma_-$    &    0 & ${\cal M}_{-}$      & -$\xi \Sigma_-$      & 0    &    -${\cal N}_{-}$ & -$\xi \Sigma_-$     \\ \hline
$\Sigma_{3}$ &  &  &     &     &     &    0 & $\xi {\cal M}_{+}$      & -$\xi {\cal M}_{-}$      & 0    &    $\xi {\cal N}_{+}$ & -$\xi {\cal N}_{-}$    &    0 & $\xi {\cal U}_{+}$      & -$\xi {\cal U}_{-}$      & 0    &    $\xi {\cal V}_{+}$ & -$\xi {\cal V}_{-}$    &    0 \\ \hline \hline
${\cal M}_{+}$ &       &       &     &    &    &     & 0      & $2 {\cal M}_{3}$      & -${\cal M}_{+}$    &    0 & 0    &    0 & 0      & -$\Sigma_+$      & -$\xi {\cal M}_{+}$    &   ${\cal Q}_{+}$ & 0    &  $\xi {\cal M}_{+}$ \\ \hline
${\cal M}_{-}$ &       &      &     &     &     &     &       & 0      & ${\cal M}_{-}$    &    0 & 0    &    0 & $\Sigma_-$      & 0      & $\xi {\cal M}_{-}$    &    0 & -${\cal Q}_{-}$    &    -$\xi {\cal M}_{-}$ \\ \hline
${\cal M}_{3}$ &  &  &     &   &     &     &       &       &    0 &    0 & 0    &    0  & $\xi {\cal U}_{+}$      & -$\xi {\cal U}_{-}$      & 0    &    -$\xi {\cal V}_{+}$ & $\xi {\cal V}_{-}$    &    0 \\ \hline
${\cal N}_{+}$ &       &       &     &     &     &     &       &       &     &    0 & $2 {\cal N}_{3}$    & -${\cal N}_{+}$ & -${\cal Q}_{+}$      & 0      & $\xi {\cal N}_{+}$    &    0 & $\Sigma_+$    &  -$\xi {\cal N}_{+}$ \\ \hline
${\cal N}_{-}$ &       &       &     &      &    &     &       &       &     &     & 0    &  ${\cal N}_{-}$ & 0      &${\cal Q}_{-}$      & -$\xi {\cal N}_{-}$   &  -$\Sigma_-$ & 0    &  $\xi {\cal N}_{-}$ \\ \hline
${\cal N}_{3}$ &  &  &     &     &     &     &       &      &     &     &    &    0 & -$\xi {\cal U}_{+}$       & $\xi {\cal U}_{-}$       & 0    &   $\xi {\cal V}_{+}$  & -$\xi {\cal V}_{-}$     &    0 \\ \hline  \hline
${\cal U}_{+}$ &       &       &     &     &     &     &       &       &     &     &     &     & 0      & $2 {\cal U}_{3}$       & -${\cal U}_{+}$     &    0 & 0    &    0 \\ \hline
${\cal U}_{-}$ &     &       &     &     &     &      &       &       &     &     &     &     &       & 0      & ${\cal U}_{-}$     &    0 & 0    &    0 \\ \hline
${\cal U}_{3}$ &  &  &    &     &     &     &       &       &    &     &   &     &       &       & 0    &    0 & 0    &    0 \\ \hline 
${\cal V}_{+}$ &      &       &     &     &    &     &       &       &    &     &    &     &       &       &     &    0 & $2{\cal V}_{3}$     &  -${\cal V}_{+}$  \\ \hline
${\cal V}_{-}$ &       &       &     &      &   &    &       &       &     &     &     &     &       &       &     &     & 0    &    ${\cal V}_{-}$  \\ \hline
${\cal V}_{3}$ & &  &   &   &   &    &       &       &     &    &    &    &       &      &    &     &     &    0 \\ \hline
\end{tabular}
\end{center}
\caption{The commutators of all $SU(4)$ superoperators.}
\end{table}

\space

\bibliographystyle{siam}

\end{document}